\def\RELEASE{1}  %
\def\ANON{0}     %
\def\SQUEEZE{0}  %

  \documentclass[sigconf,authorversion,nonacm]{acmart}

\PassOptionsToPackage{hyphens}{url}\usepackage{hyperref}

\usepackage[noend]{algpseudocode}
\usepackage{algorithmicx}
\usepackage{booktabs}
\usepackage{caption}
\usepackage{subcaption}
\usepackage{comment}
\usepackage[inline]{enumitem}
\usepackage{graphicx}
\usepackage{listings}
\usepackage{multirow}
\usepackage{xcolor}
\usepackage{xspace}
\usepackage{microtype}
\usepackage{tikz}
\usepackage{gnuplot-lua-tikz}
\usepackage{array}
\usepackage{tabularx}
\usepackage{pifont}

\microtypecontext{spacing=nonfrench}

\hypersetup{
  pdflang={en},
  pdfdisplaydoctitle}

\definecolor[named]{OurPurple}{cmyk}{0.55,1,0,0.15}
\definecolor[named]{OurDarkBlue}{cmyk}{1,0.58,0,0.21}
\hypersetup{
  colorlinks,
  linkcolor=OurPurple,
  citecolor=OurPurple,
  urlcolor=OurDarkBlue,
  filecolor=OurDarkBlue}

\if \SQUEEZE 1
\setlist[itemize]{
  leftmargin=*,
  itemsep=2pt,
  topsep=2pt}

\fi
\setcopyright{none}
\renewcommand\footnotetextcopyrightpermission[1]{} %

\def\Snospace~{\S{}}

\newcommand{\fakepara}[1]{\vspace{0mm}\noindent\textbf{#1}\quad}

\if \RELEASE 1
  \def\NOTES{0}
\else
  \def\NOTES{1}
\fi
\if \NOTES 1
  \newcommand{\XXX}[1]{{\color{red}{XXX {#1}}}}
  \newcommand{\antoine}[1]{{\color{teal}{[\textbf{AK:} {#1}]}}}
  \newcommand{\hj}[1]{{\color{violet}{[\textbf{HJ:} {#1}]}}}
  \newcommand{\lijl}[1]{{\color{orange}{[\textbf{JL:} {#1}]}}}
  \newcommand{\ja}[1]{{\color{cyan}{[\textbf{JA:} {#1}]}}}
  \newcommand{\va}[1]{{\color{olive}{[\textbf{VA:} {#1}]}}}
  \newcommand{\todo}[1]{{\color{blue}{TODO: {#1}}}}
\else
  \newcommand{\XXX}[1]{}
  \newcommand{\antoine}[1]{}
  \newcommand{\hj}[1]{}
  \newcommand{\lijl}[1]{}
  \newcommand{\ja}[1]{}
  \newcommand{\va}[1]{}
  \newcommand{\todo}[1]{}
\fi

\newcommand{\ie}{i.e.\xspace}

  \newcommand{\sys}{Columbo\xspace}

\begin{document}
\date{}
\title[\sys]{\sys: Low Level End-to-End System Traces through Modular Full-System Simulation}
\subtitle{Vision Paper (\pageref{page:last} pages total)}

  \author{Jakob Görgen}
  \affiliation{
    \institution{Max Planck Institute for Software Systems}
    \city{Saarbrücken}
    \country{Germany}}
  \email{jgoergen@mpi-sws.org}

  \author{Vaastav Anand}
  \orcid{0000-0001-8502-0657}
  \affiliation{
    \institution{Max Planck Institute for Software Systems}
    \city{Saarbrücken}
    \country{Germany}}
  \email{vaastav@mpi-sws.org}

  \author{Hejing Li}
  \orcid{0000-0001-6930-2419}
  \affiliation{
    \institution{Max Planck Institute for Software Systems}
    \city{Saarbrücken}
    \country{Germany}}
  \email{hejingli@mpi-sws.org}

  \author{Jialin Li}
  \orcid{0000-0003-3530-7662}
  \affiliation{
    \institution{National University of Singapore}
    \city{}
    \country{Singapore}}
  \email{lijl@comp.nus.edu.sg}

  \author{Antoine Kaufmann}
  \orcid{0000-0002-6355-2772}
  \affiliation{
    \institution{Max Planck Institute for Software Systems}
    \city{Saarbrücken}
    \country{Germany}}
  \email{antoinek@mpi-sws.org}
\begin{abstract}
  Fully understanding performance is a growing challenge when building
  next-generation cloud systems.
  Often these systems build on next-generation hardware, and
  evaluation in realistic physical testbeds is out of reach.
  Even when physical testbeds are available, visibility into essential
  system aspects is a challenge in modern systems where system
  performance depends on often sub-$\mu s$ interactions between HW and
  SW components.
  Existing tools such as performance counters, logging, and
  distributed tracing provide aggregate or sampled information,
  but remain insufficient for understanding individual requests
  in-depth.

  In this paper, we explore a fundamentally different approach to
  enable in-depth understanding of cloud system behavior at the
  software and hardware level, with (almost) arbitrarily fine-grained
  visibility.
  Our proposal is to run cloud systems in detailed full-system
  simulations, configure the simulators to collect detailed events
  without affecting the system, and finally assemble these events into
  end-to-end system traces that can be analyzed by existing distributed tracing tools.
\end{abstract}
 \maketitle

\section{Introduction}

Understanding performance of a modern cloud or datacenter system in
depth during development is increasingly difficult.
Modern systems are heterogeneous and require more complex and
fine-grained interaction between various software and hardware
components.
Heterogeneous systems with specialized hardware and software
components~\cite{obleukhov22precision,li2023pond,bobda2022future,xekalaki2022enabling,firoozshahian2023mtia,tal2024nextgen,abadi:tensorflow,yang2023fiber}
improves performance and efficiency, but also at the expense of understandability.

With specialized hardware, such as computational accelerators,
SmartNICs, smart SSDs, disaggregated memory, or in-memory compute,
overall system performance often critically depends on frequent,
$\mu{}s$- or $ns$-scale, and asynchronous interactions between
hardware and software.
To make matters worse, hardware and software components are typically
tightly integrated, e.g., through kernel-bypass, resulting in complex
request control flows and no common points for easy observability, such as the kernel syscall interface.
Attempts to instrument these systems for visibility into performance
either provide visibility limited to aggregated data, such as performance
counters, or incur prohibitive overheads that also significantly
change system behavior.

Building next-generation research systems frequently brings additional
challenges with hardware, often because appropriate physical testbeds
are out of reach.
Some work may explore emerging hardware or hardware features, where
hardware may be not yet available or may have substantial bugs or
limitations.
For work proposing new hardware changes, evaluation is even more
challenging, as prototyping hardware takes significantly longer and is
often prohibitively expensive.
As a result, a thorough end-to-end evaluation for in-depth performance
is either not possible, or heavily limited in insights.

Both of these problems ---
in-depth visibility and lack of physical testbeds ---
can be simultaneously addressed by simulation.
Simulations offer four key advantages:
(i) they can provide virtual testbeds for systems where a physical
testbed is unavailable.
(ii) virtually \emph{unlimited visibility into behavior of the
simulated system without affecting system behavior} through detailed
simulator logs.
Detailed simulator logs, e.g., logging each instruction or cache
line accessed by a processor, or reception and transmission time of
each packet, might substantially slow down
the simulator itself, but have no impact on the simulated system in any way.
(iii) they can provide ground-truth information hard to
obtain in physical systems, such as precise timing for events or
access low-level hardware interactions such as memory bus
transactions.
For instance, simulation provides a true and precise global clock for
all events.
(iv) they can get data from multiple components within the same execution
context to establish causal links between events originating from
different components.

As no individual simulator supports simulating all components needed
for most data center systems, modular full system
simulation~\cite{li:simbricks,rodrigues:sst}, i.e., combining and connecting
multiple existing simulators for different
components into complete simulated testbeds, is the only available
means for simulating complete cloud and datacenter systems.
Full-system simulation includes all hardware and software, including
applications, libraries, and operating system.

Can existing modular simulators readily address the visibility challenge?
Unfortunately, the answer is no.
Doing so requires solutions to the following hard issues.
First, simulators generate large amount of logging data.
As an example, a gem5 simulator~\cite{binkert:gem5} instance can
generate 100s of GBs log for a few seconds of simulated execution. Second, data from multiple simulators can be difficult to correlate.
Establishing causality of events is thus a non-trivial task.
Third, log data does not have a common format.
They can be semi-structured, and lacks any standardization across simulator types.
Lastly, there is a dearth of tooling available to analyze the data
generated by such simulations.

To address this, in this paper, we propose combining end-to-end
simulations with common monitoring techniques such as distributed
tracing~\cite{fonseca2007x,barham2004using,sigelman2010dapper}
represents a means for developers to gain visibility and
insights into heterogeneous systems. We have developed \sys, a prototype
efficiently processing the logs generated from simulations
to generate traces that can be digested and analyzed existing
distributed tracing infrastructure. We show \sys's
efficacy in analyzing heterogeneous systems by using \sys to show how
a user can gain deep insight into issues arising from external factors when using NTP for doing clock synchronization. We
believe that the approach provided by \sys to gain insight into
heterogeneous systems is effective and time-saving for developers
building heterogeneous distributed systems for the cloud.
\section{Background and Motivation}%
\label{sec:bg}

\subsection{Modular e2e Simulations}

Modular end-to-end simulators are particularly useful for
heterogeneous computer systems. Although many simulators have been
long established and used in research on  various layers of modern
systems, these simulators focus on specific components of the entire
system. Thus, none of these simulators in isolation enables a true
end-to-end evaluation of a complete modern data-center system.
For instance, computer architects use gem5~\cite{binkert2011gem5} to
explore new processors and memory modules etc., network researchers
use ns3~\cite{riley2010ns} to evaluate network protocols and
topologies, and hardware researchers use simulators like
verilator~\cite{software:verilator} for hardware RTL design.

Modular end-to-end
simulation~\cite{li:simbricks,rodrigues:sst,menard:gem5systemc}, on
the other hand, allows users to combine and connect multiple component
simulators to construct a full system simulation.
SimBricks~\cite{li:simbricks}, for example, provides fixed component
simulator interfaces to connect component simulators running as
separate processes.

While each component simulator can provide highly detailed information
about the behavior of that component without affecting system
behavior, manually analyzing and correlating these multiple logs and
reasoning about the behavior of the full system is both challenging
and extremely time-consuming. Understanding behavior across components
also requires a deep understanding of the components and their
interactions in the system. The log messages and actions triggered by
each component simulator in response to a particular event of interest
lack the context of the event. Consequently, users must manually
correlate cross-component events using detailed information such as
timestamps, DMA request IDs, and accessed memory addresses.

\sys aims to address these challenges by automatically analyzing the
correlations between events generated by different components and
visualizing them as distributed traces. By providing a comprehensive
overview of the system's behavior, \sys facilitates easier
interpretation and analysis of complex interactions within the system,
significantly reducing the effort to understand and debug the system
performance.

\subsection{Distributed Tracing}

Distributed tracing tools are widely used both in open-source~\cite{jaeger,zipkin,opentelemetry} and major internet companies~\cite{sigelman2010dapper,pandey2020building,kaldor2017canopy} to chronicle the structural execution of an end-to-end request~\cite{barham2004using,fonseca2007x}. Traditionally, the request's execution flow is captured across different components of a distributed system but the capture is limited to the application layer at each component.

The big advantage of using distributed tracing is that it is particularly useful for troubleshooting cross-component issues as it can provide insight for a failure from different components in a single context using context propagation~\cite{mace2018pivot}. A drawback of using distributed tracing is that it adds an extra penalty on the performance of the application being traced leading practitioners to use sampling techniques to effectively limit the data being traced~\cite{sigelman2010dapper,las2018weighted,las2019sifter} which can cause a potential loss of edge-case data~\cite{zhang2023benefit}. Moreover,the source data for distributed traces is currently limited to the data generated by the applications and no data from lower layers in the system software stack.

\subsection{The Case for Combining Distributed Tracing with E2E Simulation}

Using distributed tracing with simulations provides a unique opportunity to leverage the full power of distributed tracing without worrying about the potential drawbacks of using distributed tracing. As end-to-end simulation is not a performance or a time-critical task, we can run distributed tracing at full tilt to generate traces for all execution flows into the system without requiring any kind of sampling to artificially limit the number of traces. Moreover, we can enrich distributed traces with fine-grained information produced from simulators to capture detailed execution flows from different hardware components to generate hardware-enriched traces to provide higher visibility into heterogeneous systems.
\section{Design}%
\label{sec:design}

\begin{figure*}[ht]%
\centering%
\if \ANON 1%
\includegraphics[width=0.98\textwidth]{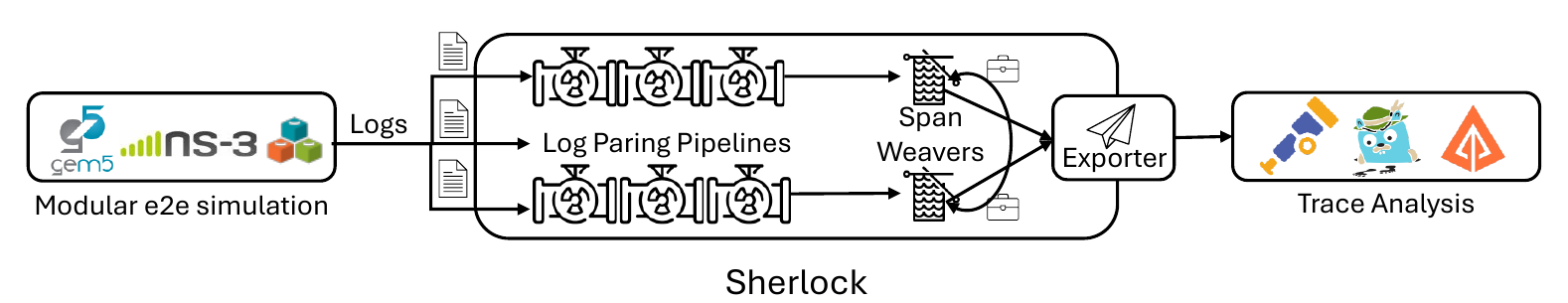}%
\else
\includegraphics[width=0.98\textwidth]{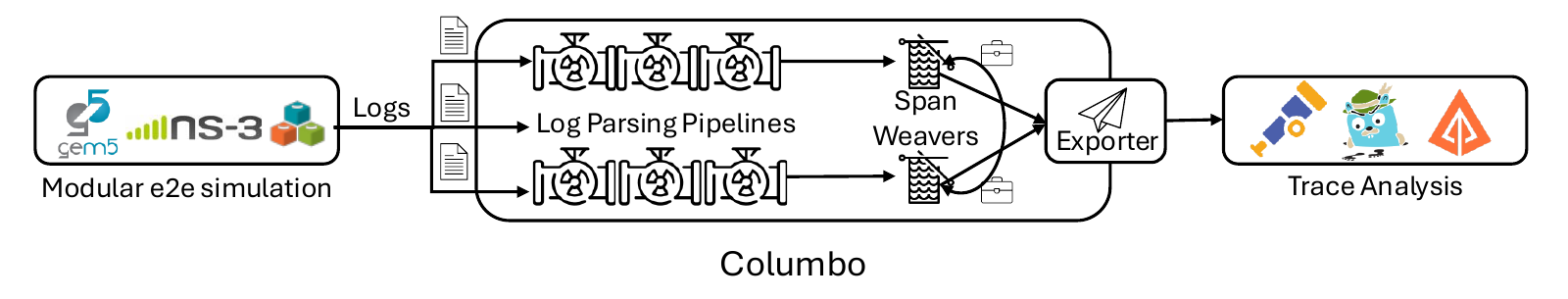}%
\fi
\caption{\sys overview}%
\label{fig:overview}%
\end{figure*}

\subsection{Design Goals}
To accurately analyze the performance of a distributed system built on a heterogeneous computing environment, we have the following design goals for ~\sys.

\begin{itemize}
    \item End-to-end visibility: Provide the full system information including host, NIC, network with the entire software stack from application to operating system.
    \item Extensibility: Allow easy integration and support of new component simulators into the tracing system.   
    \item Arbitrarily detailed information: Flexibly control the level of detail for logging based on the need.
    \item Transparent: Collect complete system log for debug and analyze without affecting distributed system performance
\end{itemize}

\subsection{Challenges}

While combining modular end-to-end simulation with distributed tracing provides a unique opportunity to gain deep insight into the system, we need to overcome many challenges to efficiently process the generated log files to produce information-rich traces that can produce actionable insight.

\fakepara{Non standardized data formats.} We are faced with the issue of non standardized
log file output formats when running modular full system simulations. Standalone
simulators for different components have different log file formats and their respective
log files contain different information. Moreover, different simulators of
the same kind log information in different formats. There is a significant lack of standardization of simulator log data which makes it difficult to easily manage and connect the simulation data for gaining insight.

\fakepara{Large amounts of data.} Simulators can provide arbitrary deep visibility
into what a simulated system does. Naturally, this visibility come at the price
of huge amounts of log file data being generated. Thus, we must be able to efficiently
handle vasts amount of data coming from multiple simulators. Moreover, a user might only be interested in a small subsection of the data for their investigation. Given the large amount of data generated, the user might not be able to easily filter out and find the relevant information.

\fakepara{Data correlation across simulator boundaries.} As we use log files
from modular, independent, unmodified simulators, making causal connections is difficult across simulator boundaries is non-trivial for two reasons. First, we cannot make any assumptions about the topology being simulated. Thus connecting events across simulator boundaries requires domain-specific knowledge of the topology as well as the simulation to be able to explicitly connect the data. Second, we cannot use any explicit context propagation techniques as we do not want to modify simulators because modifying simulators will break the modularity as well as make the system less extensible to new simulators and components.

\fakepara{Trace analysis.} Users of \sys need an easy way to analyze the traces
created. Analysis tools must support efficient querying, expressive analysis, as well as a usable interface for users to interact with the generated traces.

\subsection {Overview}

We have designed \sys to be able to process logs from modular end-to-end simulations to generate distributed traces. These traces can then be exported existing trace analysis tools for further processing and utilization by developers. \autoref{fig:overview} shows the current design architecture of \sys. Simulation logs are input to the \sys processing pipeline. To convert these logs into traces, \sys uses simulator-specific parsers to generate \emph{type-specific event streams} to generate a standard set of events for each simulator type. The generated event streams are further processed by simulator-specific \emph{pipelines} where events can be filtered, modified, or removed depending on the requirements of the user. Finally, the filtered events are passed to the \emph{SpanWeavers} which coalesce the events into spans. \emph{SpanWeavers} also do context propagation to correctly connect spans from different simulators.

\subsection{Type-Specific Event Streams}
\label{sec:eventstream}

We use modular full-system simulation consisting of multiple simulators,
each of which simulates a specific component. Each simulator creates logs in an
ad-hoc custom format with no standardization. This leads to two different standardization
challenges.

First, simulators of different types have different types of events.
For example, consider the events generated by a NIC simulator and host simulator during the simulation of a packet transmission. The host simulator might send a
packet during a simulation. For the host simulator, the generated events include function call events to invoke the NIC's driver as well as a mmio write event responsible
for instructing the attached NIC to send a packet. The events generated on the NIC side
are the dma access events generated by the NIC to read the packet data
from the host and a packet transmission event when the NIC finally puts the packet
onto the wire to send it.

Second, there is a lack of
standardization of log formats across different simulators for the same type. For
example, the logfile format of the gem5 host simulator~\cite{binkert2011gem5} is
different from the logfile format of the QEMU host simulator~\cite{software:qemu}.

To fix these issues, we introduce type-specific event streams. We create standard types
of events a simulator can emit for every given simulator type. This handles data
from two different simulators of the same type in a standardized way as each simulator of a specific type strictly adheres
to the same set of events. This eliminates the differences
between events generated from different simulator types and enables supporting new
simulators with little effort as it only needs to provide a parser for the simulator's specific logfile
format has to be provided.

\subsection{Pipelines}

\begin{figure}[h]%
\centering%
\includegraphics[width=0.48\textwidth]{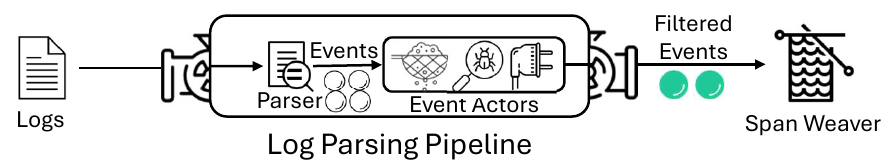}%
\caption{\sys pipelines overview}%
\label{fig:pipelines}%
\end{figure}

Depending on the granularity at which a given simulator operates, the logfiles generated by the simulator can contain large amounts of data.
For example, logfiles generated by gem5~\cite{binkert2011gem5} can have hundreds of Gbs or even TBs in size.
Additionally, not all the data generated by a simulator might be useful for the user.

To solve the problem of dealing with the large amounts of data generated, we introduce the notion of
simulator-specific pipelines. Simulator-specific pipelines are used by \sys to stream
and process the data and events contained in the logfiles generated by the simulators.
\autoref{fig:pipelines} shows an example of how multiple simulator-specific pipelines operate in conjunction together.

A pipeline consists of a producer, multiple optional actors, and a consumer. Producers read and parse the log file of a simulator and generate an event stream as described in \autoref{sec:eventstream}. Actors serve as additional operators on
the event stream within a pipeline, for example, to filter events from an event stream
or to modify the event stream to resolve a function's address to the function's
name. Consumers are the ending stage of a simulator specific pipeline and are called \emph{SpanWeavers}.
SpanWeaver group the events of an event stream into traditional distributed tracing spans and are
responsible for trace context propagation, thus creating causal relationships between spans.
An example of a simulator specific pipeline is depicted in~\autoref{fig:pipelines}.
The various components of a pipeline (i.e. producer, actors \& consumer) communicate
with each other via message queues that may be distributed over the network.

\subsection{Context Propagation}

As we use unmodified simulators for the simulation, using explicit mechanisms to propagate
trace context is infeasible as it would require non-trivial modifications to the simulators.
\sys relies on a form of implicit context propagation in which the logic of when 
to propagate context is baked into the SpanWeavers operating on a type-specific event stream.
Therefore, a span weaver instance is only responsible for the creation of 
spans and the propagation of trace context in regards to the event stream it operates on.

The choice of when to propagate the context is provided to the user as the user is responsible for configuring events from the event stream into spans. This is because the topology of the system being simulated is different for each experiment, so \sys cannot hardcode any of the context propagation a priori.
\sys's current prototype uses a simple form of context propagation in which we distinguish 
two cases.

\fakepara{Intra SpanWeaver propagation.} SpanWeavers create spans from the events 
contained in the event stream it operates on. For a specific simulator's event stream, 
a SpanWeaver creates a variety of spans, each reflecting a 
unit of work done in the respective simulator. For example, for the event streams of host simulators, the SpanWeaver might create many different spans such as a span representing a syscall, a span 
representing a mmio-write access to an attached device, among others. Spans created within 
the same SpanWeaver can have a causal relationship. For example, a syscall might trigger the host to write a register of an attached device. Therefore, context must be 
propagated between two such spans. The logic to propagate trace context between 
spans within a SpanWeaver must be generically programmed into a SpanWeaver and requires 
domain specific knowledge from the user. 

\fakepara{Inter SpanWeaver propagation.} Similar to the causal relationships 
between spans within a SpanWeaver instance, spans created by different SpanWeaver 
instances may also have causal relationship. For example, consider a span 
representing a mmio-write access to a register of an to the host attached device. 
This mmio-write will cause events in the simulator simulating the attached 
device which in turn will be grouped into a logical unit of work i.e. a span by a 
SpanWeaver. Between the two spans, created by different SpanWeaver instances, a 
causal connection must be made to preserve the relationships between different events. 
This kind of context propagation becomes necessary 
when simulators communicate through natural boundaries that also exist in real system 
like PCIe or Ethernet. Therefore, programmers must take into account when events 
cause events in attached simulators, \ie, a component would communicate over PCIe 
or Ethernet in a real system, and propagate context to another SpanWeaver instance 
in such a case. This is done through message passing queues that SpanWeaver instances 
share. In situations where a span might cause actions in a different simulator, the SpanWeaver must 
push context to that queue for the SpanWeaver of the other simulator. Vice versa, in case a span might have been caused by an 
attached simulator a SpanWeaver must poll context from that queue in order to make 
the causal connection.

To make it easy for users to do context propagation, we limit the number of different types of spans that can be generated from an event span. This allows users to programmatically encode the causal relationships between spans.

\subsection{Exporter}

Once a SpanWeaver has finished a span and made all the necessary causal connections,
it passes the span to an exporter. Exporters are used by \sys to export spans to
external tools from the distributed tracing community, such as Jaeger~\cite{jaeger} or OpenZipkin~\cite{zipkin}.
These tools provide \sys users with a graphical representation of the created spans
and traces. Since such tools may use a different internal representation of spans
and traces than \sys, the exporters are also responsible for converting \sys's internal
span representation into the representation required by the tool a user wants to
export traces to, before actually exporting/sending traces to the respective tool.

\subsection{Online Analysis}

It is desirable to run \sys in parallel to the actual full-system simulation without
the need to persist the log files written by the simulators for analysis after the
simulation finished. This is necessary as users might not have hundreds of GBs of
disk space at their disposal to persist logfiles for offline analysis.

Therefore, \sys provides an online mode with the support of Linux named pipes.
Each simulator in the simulation uses a Linux named pipe as the destination for it's log events.
\sys is started in parallel to the simulation to read from the named pipes to create traces. This allows \sys to create traces without requiring any persistence of the log data.
\section{Implementation}%

\begin{table}[]
    \begin{tabular}{|m{0.13\textwidth}|m{0.13\textwidth}|m{0.13\textwidth}|}
    \hline
    \textbf{Simulator Type} & \textbf{\# Supported Event Types} & \textbf{\# Supported Span Types} \\ \hline
    Host                    & 16                                & 6                                \\ \hline
    NIC                     & 9                                 & 4                                \\ \hline
    Network                 & 3                                 & 1                                \\ \hline
    \end{tabular}
    \vspace{.3cm}
    \caption{Supported simulator types and the number of supported events and span types respectively.}%
    \label{fig:amount-types}%
\end{table}

The current \sys prototype is completely implemented in C++ and is tailored 
for use together with SimBricks~\cite{li:simbricks} simulations. 
It consists of 19,000 lines of code and supports three simulators: the gem5 simulator~\cite{binkert2011gem5} to simulate end-hosts, the ns3
network simulator~\cite{riley2010ns}
as well as the SimBricks Intel i40e NIC behavioral model simulator~\cite{li:simbricks}.
The amount of supported event- and span-types for each of these simulators is listed in~\autoref{fig:amount-types}.

\fakepara{\sys scripts.}\sys harnesses logfiles written by simulators during a modular full system simulation.
As \sys is decoupled from the actual simulation and the simulators used, \sys does 
not automatically know about the simulated topology and the concrete simulator instances used 
for the components involved. Therefore, a user has to orchestrate the trace 
creation by providing a \sys \emph{Script}. A \sys \emph{Script} is a small C++ program 
with the purpose of composing simulator specific pipelines similar to ns3~\cite{riley2010ns} requiring 
users to write small C++ programs in order to create experiments. To make it easy for users to write \sys \emph{Scripts}, \sys 
provides predefined building blocks like logfile parsers, event stream actors, 
SpanWeavers, and predefined components to export the traces created to external 
tools for visualization and analysis. Using these building blocks users can easily 
compose simulator specific pipelines in order to created low level end-to-end system 
traces trough simulations.
\section{Case Study: Clock Synchronization}%
\label{sec:eval}

In this case study, we consider the scenarios where we want to synchronize the clocks in our system.
To do so, we need to synchronize the clock of the client with the clock of the server.
The client and server are connected over the network which is also servicing background traffic generated from other systems and applications.
Checking if distributed clocks are synchronized is a difficult problem as there is no ground truth available and it is impossible to guarantee that the measurements taken at different clocks will be done so at the exact moment of time.

\begin{figure}[h]%
\centering%
\includegraphics[width=0.48\textwidth]{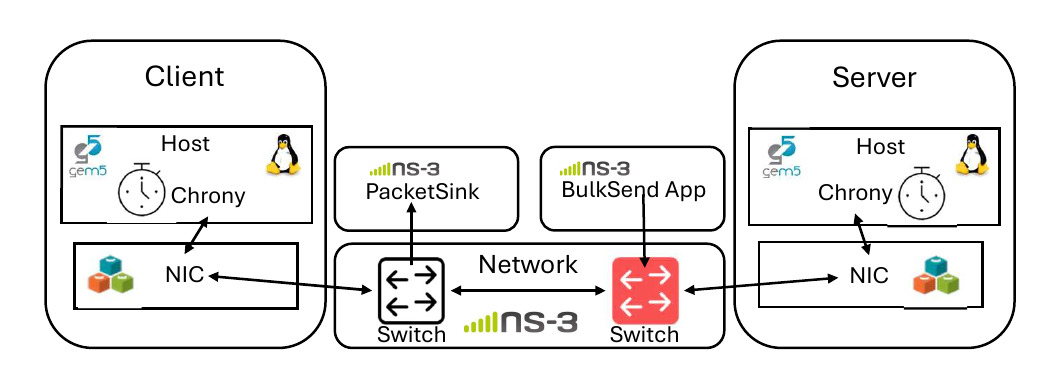}%
\caption{\sys evaluation topology}%
\label{fig:eval-topo}%
\end{figure}

Instead, to check if the clocks are synchronized, we turn to simulation where we simulate the client, the server, and the underlying network as shown in \autoref{fig:eval-topo}. We simulate the topology with SimBricks~\cite{li:simbricks}. The topology 
consists of two hosts, the client and server, each connected to a NIC, connected over a network 
consisting of two switches. To simulate the two hosts, we used the gem5 simulator~\cite{binkert2011gem5}, 
to simulate the network we used ns3~\cite{riley2010ns} and to simulate the two NICs 
we used SimBricks's Intel i40e NIC behavioral model simulator~\cite{li:simbricks}.

To synchronize the clocks we run chrony~\cite{chrony} on both the client and server. We configure chrony to use NTP for doing clock synchronization.
We check if the clocks can be synchronized under two different conditions. In \emph{Scenario 1}, we check if the clients can be synchronized when there is no background traffic from other applications. This will help us establish a baseline of how well the clocks can be synchronized in best conditions. In \emph{Scenario 2}, we check if the clients can be synchronized when there is a lot of background traffic from other applications. To simulate the background traffic, we use the BulkSendApplication and PacketSink applications from NS3 to send TCP packets from the application to the sink. We configured the BulkSendApplication to transmit packets at a data rate that exhausts the link between the two switches.

We ran all experiments on a physical host with two Intel(R) 
Xeon(R) Gold 6336Y CPU @ 2.40GHz, 24 cores each, with 8x 32GiB
(256GiB) of memory. 
During all experiments we enabled log output for the simulators used, i.e. 
we enabled logging statements when compiling the Intel i40e NIC behavioral model, 
we used and attached ns3 trace sources, and we used gem5 in the \emph{opt} version 
together with debugging flags we passed via the command line.

\begin{figure}%
\centering%
\includegraphics[width=0.48\textwidth]{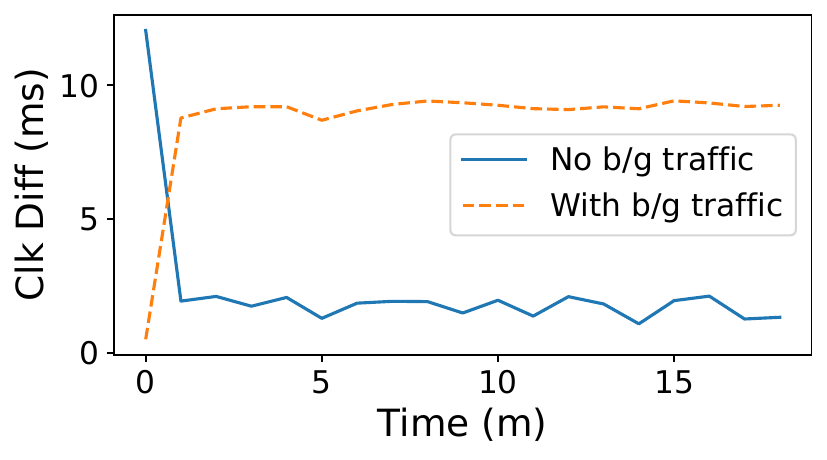}%
\caption{Measured difference between the system clocks of the client and server}%
\label{fig:sys_clk_diff}%
\end{figure}

We measure the difference between the client and server system clocks.
To ensure that we take a clock reading at the same time, we make use of the global clock available to us in the simulation.
Each clock reading has a corresponding timestamp with respect to the global clock. This allows us to apply corrections and ensure
that we are reading the clocks at the same time.
\autoref{fig:sys_clk_diff} shows the measured difference in the clocks for both scenarios. We notice that the system does a good job of synchronizing the clocks in the scenario where there is no background traffic. However, in the scenario with background traffic, the clock synchronization does not achieve the same accuracy as the baseline.

\begin{figure}%
\centering%
\includegraphics[width=0.48\textwidth]{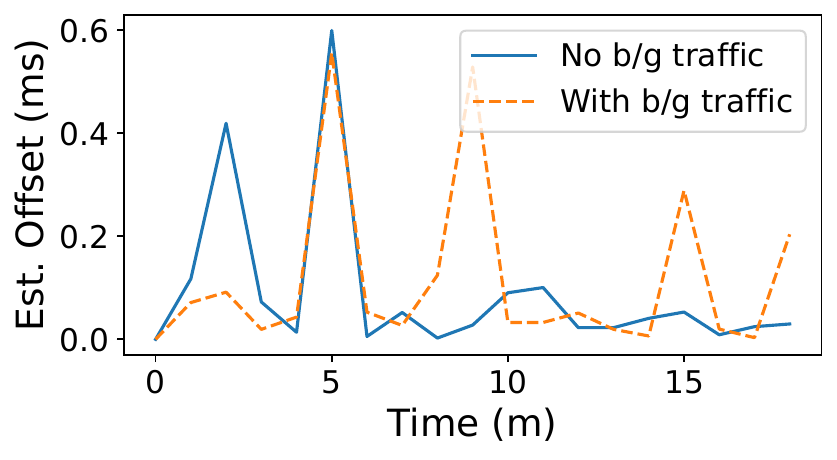}%
\caption{Clock offset between client and server clocks estimated by chrony}%
\label{fig:chrony_est}%
\end{figure}

To figure out why the client and server clocks are not synchronized when there is background traffic, we investigate further and look into the clock offset calculated by chrony. \autoref{fig:chrony_est} shows the estimated clock offsets for both the scenarios. For both scenarios, the clock offsets have almost identical trends but we know that this is not accurate as our system clock measurement is still not accurate. Thus, we can conclude that there is something odd happening during our clock synchronization but at the moment we do not have enough information to figure out the odd behavior.

\begin{figure}%
\centering%
\includegraphics[width=0.48\textwidth]{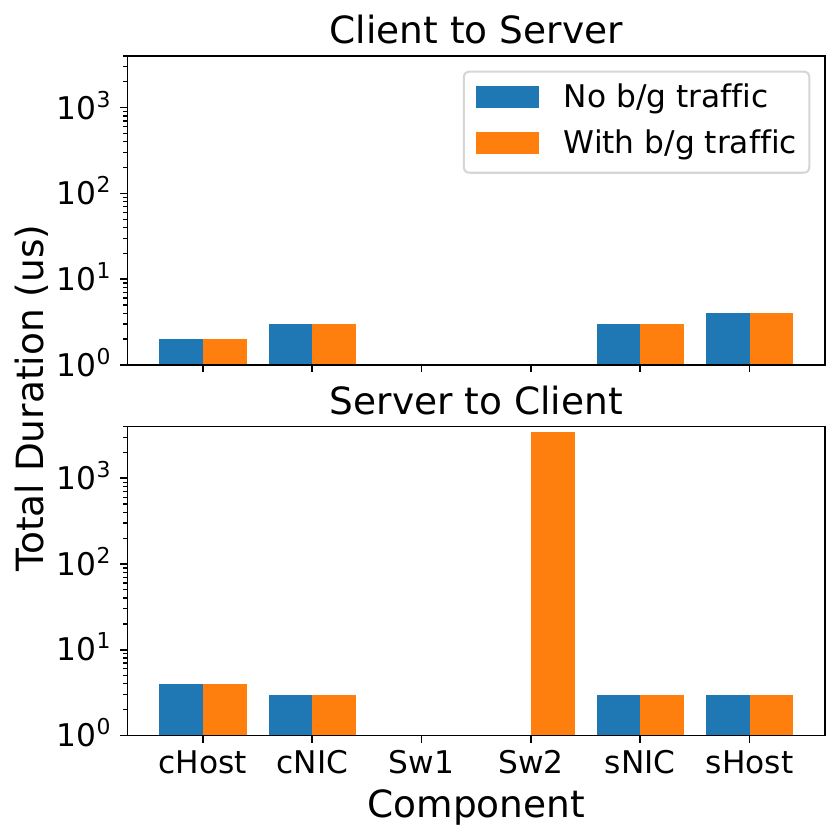}%
\caption{Breakdown of the time spent per component for clock synchronization request}%
\label{fig:visibility}
\end{figure}

To figure out the odd behavior, we need to dive deeper into the clock synchronization process between the server and client. \autoref{fig:visibility} shows the total time spent per component for the two different scenarios. We break it down into two different parts - (i) the synchronization request sent from the client to the server, (ii) the synchronization response sent from the server to the client. As we can see that the big difference in the two situations is that the response spends considerably large amount of time in the switch that is connected directly to the BulkSendApplication. Thus, as the request and response messages take different durations, this makes NTP inaccurate as one of the requirements for NTP is that both the request and response should take a similar amount of time. The added visibility allows us to easily figure out the root cause of the inaccuracy in our synchronization.

\section{Challenges \& Opportunities}
\label{sec:discussion}

\fakepara{Correct Context Propagation:} Currently, \sys infers causal
relationships between events without explicit context propagation.
This is error-prone in scenarios with multiple parallel events triggered by different
components. This error could potentially be prevented by
explicit context propagation, but it requires a significant and error prone implementation effort. Moreover, this approach would undermine \sys goal of allowing for easy integration of new simulators. Thus, correctly doing context propagation in all scenarios remains a significant future challenge.

\fakepara{Gaining insight into deployed systems:} Deep insight into
deployed systems remain impractical due to overhead. The emergence of
heterogeneous hardware systems with novel hardware components further
complicates this. As components are developed independently, there is
little standardization of the granularity of metrics, logs, or traces available from these components. Operators rely on bespoke solutions for gaining insight into these systems~\cite{andrei2024aiobservability}.

We believe  \sys provides an opportunity to solve in-production bugs by connecting real executions to simulation. Production systems do not offer the opportunity to gain deep insight into the system as that would drastically impact the performance. However, data collected in production could be used in simulation with \sys to gain the required visibility into existing systems and potentially solve bugs. We believe that this connection represents an exciting avenue for research in heterogeneous cloud systems.

\fakepara{Tracking down corner-cases: } Tracking down rare abnormal
behavior in distributed systems is often challenging with existing
tracing tools. The root cause is often not solely caused events
captured in traces, but also heavily depends on the overall
hardware and software state. \sys builds on typically deterministic
simulations, thus users can easily reproduce the problematic
behavior of the system with more detailed logging. We expect this
significantly enhances debugging efficiency.

\section{Conclusion}

This paper presents a novel approach to gain in-depth understanding of cloud system 
behavior. We propose leveraging detailed full-system simulations to capture (almost) 
arbitrarily fine-grained events across software and hardware layers. \sys seamlessly 
integrates with any existing simulator generating log files, requiring 
no changes to the simulators themselves. Since most existing simulators already 
provide detailed logs, no further instrumentation is needed. Instead, \sys requires 
users to create a parser specific to each simulator's log format to translate those 
logs into a type-specific event stream \sys can understand. These event streams 
are then assembled into end-to-end system traces, compatible with existing distributed 
tracing tools for analysis. 

The presented approach offers unique advantages. Firstly, simulations eliminate the performance 
concerns associated with real-world distributed tracing. Secondly, simulations allow 
us to enrich these traces with detailed information from different hardware components,
creating hardware-enriched traces for unparalleled visibility into heterogeneous systems.
By combining the power of distributed tracing with the control offered by simulations, 
\sys empowers us to gain a deeper understanding of systems, facilitating efficient
performance analysis and optimization.

\bibliographystyle{plain}
\bibliography{paper,bibdb/papers,bibdb/strings,bibdb/defs}

\label{page:last}
\end{document}